\documentclass[aps,preprint,showpacs]{revtex4}
\usepackage{graphicx}
\usepackage{bm}
\newcommand{\te}{\mathsf{t}}
\newcommand{\n}{\mathbf{n}}
\newcommand{\rb}{\mathbf{r}}
\newcommand{\btau}{\bm{\tau}}
\newcommand{\tot}{\text{tot}}
\newcommand{\hp}{\text{hp}}
\newcommand{\el}{\text{el}}
\newcommand{\vis}{\text{vis}}
\newcommand{\elm}{\text{em}}
\newcommand{\ph}{\text{ph}}
\newcommand{\Vo}{\mathcal{V}}
\newcommand{\E}{\mathbf{E}}
\newcommand{\Om}{\bm{\Omega}}
\newcommand{\Mb}{\mathbf{M}}
\newcommand{\Tb}{\mathbf{T}}
\begin{document}
\title{Light-induced rotation of dye-doped liquid crystal droplets}
\author{C. Manzo}
\affiliation{INFM-CNR Coherentia and Universit\`{a} ``Federico II'',
Dipartimento di Scienze Fisiche, Complesso di Monte S.Angelo, via
Cintia, 80126 Napoli, Italy}
\author{D. Paparo}
\affiliation{INFM-CNR Coherentia and Universit\`{a} ``Federico II'',
Dipartimento di Scienze Fisiche, Complesso di Monte S.Angelo, via
Cintia, 80126 Napoli, Italy}
\author{L. Marrucci}
\email{lorenzo.marrucci@na.infn.it}
\affiliation{INFM-CNR Coherentia
and Universit\`{a} ``Federico II'', Dipartimento di Scienze Fisiche,
Complesso di Monte S.Angelo, via Cintia, 80126 Napoli, Italy}
\author{I. J\'{a}nossy}
\affiliation{Research Institute for Solid State Physics and Optics,
H-1525 Budapest, Hungary}
\date{\today}
\begin{abstract}
We investigate both theoretically and experimentally the rotational
dynamics of micrometric droplets of dye-doped and pure liquid
crystal induced by circularly and elliptically polarized laser
light. The droplets are dispersed in water and trapped in the focus
of the laser beam. Since the optical torque acting on the molecular
director is known to be strongly enhanced in light-absorbing
dye-doped materials, the question arises whether a similar
enhancement takes place also for the overall optical torque acting
on the whole droplets. We searched for such enhancement by measuring
and comparing the rotation speed of dye-doped droplets induced by a
laser beam having a wavelength either inside or outside the dye
absorption band, and also comparing it with the rotation of pure
liquid crystal droplets. No enhancement was found, confirming that
photoinduced dye effects are only associated with an internal
exchange of angular momentum between orientational and translational
degrees of freedom of matter. Our result provides also the first
direct experimental proof of the existence of a photoinduced stress
tensor in the illuminated dye-doped liquid crystal. Finally,
peculiar photoinduced dynamical effects are predicted to occur in
droplets in which the molecular director is not rigidly locked to
the flow, but so far they could not be observed.
\end{abstract}
\maketitle
\section{Introduction}
It has been known since the early eighties that light can transfer
its angular momentum to liquid crystals with high efficiency,
causing the rotation of the average molecular orientation, as
specified by the molecular director \cite{review}.

In the early 90's it was discovered that adding small amounts of
certain dyes to the liquid crystals, the light-induced torque on the
molecular director could be greatly enhanced
\cite{janossy90,janossy92}. This effect, initially rather puzzling,
was later explained by a model relying on the reversible changes of
intermolecular forces occurring between photoexcited dye molecules and
liquid crystal host \cite{janossy94}. This model was subsequently
extended and refined \cite{marrucci97pre} and confirmed with several
independent experiments
\cite{marrucci97jcp,marrucci00,kreuzer02,kreuzer03,truong05}.

The understanding achieved with this model has also provided an
answer to a fundamental question related to the observed torque
enhancement: where does the additional angular momentum come from?
Clearly, the angular momentum that is transported by the optical
wave impinging on the material cannot be affected by the presence of
dye. It is true that absorption of light due to the dye does lead to
some additional transfer of angular momentum from light to matter
and therefore to a variation of the light-induced torque. However,
this effect cannot account for the magnitude of the observed torque
enhancement \cite{marrucci97jcp} and for its peculiar dye-structure
dependence \cite{marrucci00}. The clearest solution to this puzzle
was put forward in Refs.\ \cite{marrucci97pre} and \cite{palffy98}
and is as follows. The angular momentum transfer from light is
indeed approximately unchanged. Light absorption however triggers a
transfer of angular momentum between different \textit{internal
degrees of freedom} of the liquid crystal, namely from the
center-of-mass molecular degrees of freedom, corresponding
macroscopically to fluid flow, to the molecular-orientation degrees
of freedom, corresponding to the molecular director. Being an
internal transfer, the \textit{total} angular momentum given to the
material as a whole is not affected by the presence of dye. Another
consequence of this internal transfer is that, by the force
action-reaction principle, in dye-doped liquid crystals there should
be another (opposite) torque acting on the fluid in the form of a
photoinduced stress-tensor \cite{marrucci97pre}.

However plausible they may be, both these predictions have never
been tested directly in an experiment. Here we address these
questions by studying the rotational behavior of pure and dye-doped
liquid crystal droplets of micrometric size, optically trapped in
water, under the effect of circularly and elliptically polarized
infrared and visible laser light. In particular, we measured the
rotation speed of dye-doped droplets illuminated with visible light,
having a wavelength in the dye absorption domain, and compared it
with that of undoped droplets and with the case in which the
illumination is by infrared light, not absorbed by the dye. Should
we observe some significant dye-induced enhancement (or other
anomalies) of the droplet rotation speed, this would imply that the
current understanding of the photoinduced torque as entirely due to
an internal exchange of angular momentum is not correct, or at least
not complete.

But what do we exactly mean by droplet rotation? From a theoretical
point of view, a liquid crystal droplet actually has two independent
rotational degrees of freedom: the average molecular director,
corresponding to the average orientation of the liquid crystal
molecules in the droplet, and the internal rotational flow,
corresponding to a fluid-like motion of the molecule centers of mass
(this is already a strong simplification: a liquid crystal droplet
has actually an infinite number of rotational degrees of freedom
associated with changing inhomogeneous configurations of director
and flow). These two rotational degrees of freedom of the droplet
are always strongly interacting with each other, via viscosity,
elastic forces, and possibly also photoinduced effects. This
interaction may be strong enough to lock the two degrees of freedom
together into a single one, so that the droplet effectively behaves
as a rigid body. This is what we actually observed in our
experiments. Nonetheless, by properly choosing the experimental
parameters, it should be possible to reach a regime in which the two
rotational degrees of freedom become effectively unlocked. We would
expect that in this ``unlocked'' regime, photoinduced dye effects
become important, even if the angular momentum exchange is entirely
internal. We investigated theoretically this case and predicted the
occurrence of highly non-trivial photoinduced rotational effects,
which we then tried to observe experimentally.

This article is organized as follows. The general continuum theory of
the light-induced dynamics of dye-doped liquid crystals is reviewed in
Sec.\ II. In Sec.\ III, we then apply this theory to the case of a
spherical droplet immersed in water. With the help of several
strongly-simplifying approximations, analytical solutions are found,
predicting what should be the light-induced rotational behavior of the
droplets as a function of light power and polarization and of droplet
size. Our experiments are then described and discussed in Sec.\ IV.
Our results are finally summarized in Sec.\ V.

\section{General theory of photoinduced dynamics of nematics}
\subsection{Dynamical fields and equations}
In general, the dynamics of a nematic liquid crystal under the action
of laser light is defined by the temporal and spatial dependence of
the following fields: (i) the molecular director $\n$ specifying the
local average molecular alignment \cite{footnote1}, (ii) the velocity
vector $\mathbf{v}$ defining the flow of matter, and (iii) the
electric and magnetic fields of the optical wave, $\E$ and
$\mathbf{B}$, respectively.

The optical fields $\E$ and $\mathbf{B}$ are governed by the usual
electromagnetic Maxwell equations in (anisotropic) dielectric media
\cite{landau}. The two material fields $\n$ and $\mathbf{v}$ are
respectively governed by the director torque balance equation,
\begin{equation}
I_n\frac{d}{dt}\left(\n{\times}\frac{d\n}{dt}\right)=\btau^{\tot}
\label{eqforn}
\end{equation}
and by the Newton equation for the acceleration (or momentum
conservation law),
\begin{equation}
\rho\frac{dv_i}{dt}=\partial_j \te^{\tot}_{ji},
\label{newton}
\end{equation}
where $\partial_j$ stands for the partial derivative
$\partial/\partial{x_j}$ and the usual sum convention over repeated
indices is understood. In these equations, $\btau^{\tot}$ is the
total torque density acting on the molecular director, $\te^{\tot}$
is the total stress tensor associated with a fluid displacement with
no director rotation (this corresponds to Ericksen's definition of
stress tensors in liquid crystals \cite{degennes2ed}), $I_n$ is a
moment of inertia per unit volume associated with the rotation of
the nematic director, which is actually negligible in all practical
cases (it is included only for making the equation physical meaning
clearer), and $\rho$ is the mass density, hereafter assumed to be
constant (incompressible fluid approximation). Moreover, in both
equations above we have used the so-called ``material'' or
``convective'' time derivative, defined as
$d/dt=\partial/\partial{t}+v_j\partial_j$, corresponding to a
derivative taken following the fluid element along its motion.
Incompressibility actually adds the following constraint on the
velocity field $\mathbf{v}$:
\begin{equation}
\partial_iv_i = 0,
\label{continuity}
\end{equation}
which is a particular case of the mass continuity equation.

\subsection{Constitutive equations}
Equations (\ref{eqforn}) and (\ref{newton}) are to be completed with
the appropriate constitutive equations for the total torque density
and stress tensor. To first order in all deviations from equilibrium
and in all gradients, we may distinguish five additive and
independent contributions to both the torque density and the
stress-tensor \cite{degennes2ed,chandrasekhar2ed}: hydrostatic
pressure (hp), elastic (el), viscous (vis), electromagnetic (em),
and photoinduced (ph), the latter being that associated with dye
effects. In summary, one may write
\begin{equation}
\btau^{\tot} = \sum_{\alpha}\btau^{\alpha}
\label{torque}
\end{equation}
and
\begin{equation}
\te^{\tot}_{ji}=\sum_{\alpha}\te^{\alpha}_{ji}.
\label{stress}
\end{equation}
Each of the ten $\btau^{\alpha}$ and $\te^{\alpha}$
($\alpha=\hp,\el,\vis,\elm,\ph$) terms appearing in these two
equations has a well defined (first-order) constitutive dependence on
the dynamical fields, which we will now briefly discuss.

First, there is actually no torque density associated to pressure
effects, i.e., $\btau^{\hp}=0$ identically. The hp stress tensor
term has instead the usual simple form
$\te^{\hp}_{ji}=-p\delta_{ji}$, where $p$ is the pressure field and
$\delta_{ij}$ is the Kronecker delta. In the incompressible-fluid
approximation we are adopting, $p$ must be treated as a pure
``constraint force'', i.e., assuming just the space and time
dependence needed to ensure continuous validity of Eq.\
(\ref{continuity}).

Next, the constitutive laws of the elastic and viscous torque
densities and stress-tensors are fully standard and we refer to
Refs.\ \cite{degennes2ed} or \cite{chandrasekhar2ed} for their
explicit expressions.

Let us now turn to the electromagnetic terms. They are also standard,
but it is nevertheless convenient to introduce them explicitly here.
We assume for the time being that the electromagnetic fields present
in our systems are associated only with an approximately monochromatic
optical wave having a given vacuum wavelength $\lambda$. Let us first
introduce the uniaxial optical dielectric tensor of the liquid crystal
\begin{equation}
\varepsilon_{ij} = \varepsilon_0(\varepsilon_{\perp}\delta_{ij}+
\varepsilon_a n_i n_j),
\end{equation}
where $\varepsilon_0$ is the vacuum dielectric constant,
$\varepsilon_{\perp}$ the relative dielectric constant for
$\E\perp\n$ and $\varepsilon_a$ the relative dielectric anisotropy.
Neglecting all magnetic effects at optical frequencies, the
electromagnetic torque density is then given by
\begin{equation}
\btau^{\elm}=\langle \mathbf{D}{\times}\E \rangle
=\varepsilon_0\varepsilon_a\langle (\n{\cdot}\E)(\n{\times}\E)\rangle,
\label{tauem}
\end{equation}
where $\mathbf{D}$ is the usual dielectric displacement field
($D_i=\varepsilon_{ij}E_j$) and where $\langle\rangle$ denotes a time
average over an optical cycle. The electromagnetic stress tensor can
be written in the following form \cite{landau,footnote2}:
\begin{equation}
\te^{\elm}_{ji}=\left\langle\left(\tilde{F}_{\elm}-\rho\frac{\partial\tilde{F}_{\elm}}
{\partial\rho}\right)\delta_{ji}+D_jE_i+B_jB_i/\mu_0\right\rangle,
\label{emstress}
\end{equation}
where $\mu_0$ is the vacuum magnetic permittivity constant and
$\tilde{F}_{\elm}$ is the electromagnetic free energy density at given
electric field $\E$ and magnetic field strength
$\mathbf{H}=\mathbf{B}/\mu_0$, which in our case can be written as
\begin{eqnarray}
\tilde{F}_{\elm}&=&-\frac{1}{2}\varepsilon_{ij}E_iE_j-\frac{B^2}{2\mu_0}\nonumber\\
&=&-\frac{\varepsilon_0}{2}\left(\varepsilon_{\perp}\delta_{ij}+
\varepsilon_an_in_j\right)E_iE_j-\frac{B^2}{2\mu_0}.
\end{eqnarray}

Let us finally consider the photoinduced terms, i.e., those
appearing in dye-doped liquid crystals when illuminated with light
having a wavelength falling within the dye absorption band. These
are absolutely nonstandard. In the limit of small light intensities,
simple symmetry arguments show that the photoinduced torque density
must be identical to the electromagnetic one except for the
replacement of the dielectric anisotropy $\epsilon_a$ with a new
material constant $\zeta$, proportional to the absorbance (or to dye
concentration) \cite{marrucci97pre}. Therefore, its explicit
expression can be written as follows:
\begin{equation}
\btau^{\ph}=\zeta\varepsilon_0\langle(\n{\cdot}\E)(\n{\times}\E)\rangle
\label{tauph}
\end{equation}
(the constant $\varepsilon_0$ is inserted for making $\zeta$
dimensionless). Similarly, we may apply symmetry arguments for
identifying the most general possible expression of the photoinduced
stress tensor. This expression contains seven new unknown material
constants (all proportional to the absorbance):
\begin{eqnarray}
\te^{\ph}_{ij}\!\!\!&=&\!\!\!\left\langle\left[a_1E^2+\!a_2(\n{\cdot}\E)^2\right]\delta_{ij}+\!
a_3E_iE_j+\!\left[a_4E^2 \right.\right. \nonumber\\ && \left. \left.
\!\!\!\!\!\!\!\!\!\!+a_5(\n{\cdot}\E)^2\right]n_in_j+\!
(\n{\cdot}\E)(a_6E_in_j+\!a_7E_jn_i)\right\rangle \!\!. \label{tph}
\end{eqnarray}
The effects of this stress tensor have never been measured or even
detected in ordinary liquid crystals, although related photoinduced
effects may have been observed in polymeric nematic elastomers
\cite{finkelmann01,yu03}.

We have thus completed the set of constitutive equations needed to
close the dynamical equations (\ref{eqforn}) and (\ref{newton}).

\subsection{Angular momentum conservation}
Before moving on to the specific case of a droplet of liquid crystal
immersed in water, it is convenient to see how the law of angular
momentum conservation enters our problem.

In contrast to the case of the conservation of (linear) momentum,
which provides an additional dynamical equation, the law of angular
momentum conservation actually sets only a general constraint on the
possible constitutive laws of torque densities and stress tensors.
For any given volume $\Vo$ of material, the corresponding angular
momentum rate of change will be given by the following law:
\begin{equation}
\frac{d\mathbf{L}}{dt}=\Mb^{\tot},
\label{cardinal}
\end{equation}
where $\mathbf{L}$ is the total angular momentum within volume $\Vo$
and $\Mb^{\tot}$ is the total external torque acting on it. The two
sides of Eq.\ (\ref{cardinal}) can be deduced by multiplying the
corresponding sides of Eq.\ (\ref{newton}) vectorially by $\rb$,
integrating them over the volume $\Vo$, and then adding to them the
volume integral of the corresponding sides of Eq.\ (\ref{eqforn}).
In this way we obtain
\begin{equation}
\mathbf{L}=\int_{\Vo}(\rho\rb{\times}\mathbf{v}+I_n\n{\times}\frac{d\n}{dt})dV
\label{momang}
\end{equation}
and $\Mb^{\tot}=\sum_{\alpha}\Mb^{\alpha}$, with
\begin{equation}
M^{\alpha}_i=\int_{\Vo}(\epsilon_{ijh}x_j\partial_k\te^{\alpha}_{kh}
+\btau^{\alpha}_i)dV,
\label{couplevol}
\end{equation}
where $\epsilon_{ijh}$ is the fully antisymmetric Levi-Civita tensor
\cite{footnote3}. However, in order for Eq.~(\ref{cardinal}) to be
equivalent to a local conservation (continuity) law, it should be
possible to reduce the total external torque to a pure surface
integral over the boundary $\partial\Vo$ of $\Vo$, such as the
following:
\begin{equation}
M^{\alpha}_i=\oint_{\partial\Vo}\epsilon_{ijh}(x_j\te^{\alpha}_{kh}+
n_j\mathsf{s}^{\alpha}_{kh})dA_k,
\label{couplesurf}
\end{equation}
where $\mathsf{s}^{\alpha}_{kh}$ is a new material tensor expressing
the torque per unit area exchanged by the director $\n$ directly
through the surface and $dA_k$ denotes a vector having direction
equal to the local surface normal (pointing outward) and modulus
equal to the area of the surface element
\cite{degennes2ed,chandrasekhar2ed}.

By equating the two expressions (\ref{couplevol}) and
(\ref{couplesurf}) of the external torque for any possible volume
$\Vo$, and exploiting the standard divergence theorems, one obtains
the local identity
\begin{equation}
\btau^{\alpha}_i=\epsilon_{ijh}\te^{\alpha}_{jh}+\partial_k
(\epsilon_{ijh}n_j\mathsf{s}^{\alpha}_{kh}).
\label{tautrel}
\end{equation}
Generally speaking, this identity holds true only for $\alpha=\tot$,
and not separately for each term $\alpha=\hp,\el,\vis,\elm,\ph$.
However, since in our first-order theory these five terms can be tuned
independently from each other, identity (\ref{tautrel}) must hold true
also for $\alpha=\hp,\el,\vis,\elm,\ph$, separately.

Moreover, since in the first order approximation only the elastic
forces (torque density and stress-tensor) are taken to depend on the
director spatial gradients, we may deduce from Eq.\ (\ref{tautrel})
that $\mathsf{s}^{\alpha}_{kh}$ is nonzero only for the elastic
contribution $\alpha=\el$. In all other cases one must have
$\mathsf{s}^{\alpha}=0$ within first order approximation. Therefore
in these cases, owing to angular momentum conservation, the stress
tensor defines completely the torque density, or conversely, the
torque density defines the antisymmetric part of the stress tensor.
In particular, Eq.\ (\ref{tautrel}) with $\alpha=\ph$ yields the
following relationship between the material constants appearing in
expressions (\ref{tauph}) and (\ref{tph}):
\begin{equation}
\zeta\varepsilon_0 = a_7 - a_6.
\end{equation}

In concluding this Section, we note that the theory we have just
described and in particular the constitutive equations we have
adopted for the photoinduced torque density and stress tensor
already imply that the photoinduced angular momentum transfer
associated with the dye is fully internal to the liquid crystal.
Indeed, the flux of angular momentum through any boundary surface,
as given by Eq.\ (\ref{couplesurf}), will vanish identically if the
stress tensor vanishes on it. Since we have assumed that all
material constants $a_i$ are proportional to the dye concentration,
they should vanish on a surface lying just outside the liquid
crystal (in water) and therefore the flux of angular momentum
through such a surface will vanish, i.e., one has $\Mb^{\ph}=0$. We
note that the same argument does not hold for $\Mb^{\elm}$, as
$\te^{\elm}$ has a finite value also in the isotropic liquid.

Is there a possible way out of this conclusion, justifying perhaps a
hypothetical photoinduced flow of angular momentum out of the liquid
crystal? Within a first order theory of the constitutive equations
the answer is no. However, it cannot be a priori excluded that
higher-order terms in the constitutive equations become important in
specific situations and justify a strong exchange of angular
momentum with the outside. For example, at the surfaces between the
liquid crystal and the surrounding medium, the mass and composition
densities suffer sharp discontinuities. Therefore, a first-order
theory in the spatial gradients is clearly not justified anymore
(this, by the way, is just how surface anchoring enters the
problem). It would then be conceivable that higher-order terms in
the photoinduced torque density and stress-tensor expressions could
give rise to photoinduced surface effects leading to a significant
angular momentum exchange with the outside, i.e., to a
$\Mb^{\ph}\neq0$. In the end, this hypothesis can only be tested,
and eventually ruled out, experimentally.

\section{Droplet rotational dynamics}
Equations (\ref{eqforn}), (\ref{newton}), and Eq.\
(\ref{continuity}), supplemented with all constitutive equations for
torque densities and stress tensors, completely define the
light-induced dynamics of the liquid crystal. In the case of a
droplet of liquid crystal immersed in water one should also include
in the system the appropriate boundary conditions at the droplet
surface. We limit ourselves to mentioning them: continuity of fluid
velocity and forces across the boundary, continuity of tangential
components of $\E$ and of normal components of $\mathbf{D}$,
continuity of $\mathbf{B}$, and appropriate anchoring conditions on
$\n$. Moreover, one should account for the dynamics of the water
surrounding the droplet. The latter is also governed by Eq.\
(\ref{newton}), but with a simpler expression of the stress-tensor,
including only hydrostatic pressure, Newtonian viscosity and the
electromagnetic stress tensor.

The resulting system of equations is clearly very complex and an exact
solution can be determined only numerically. In the following, we
instead approach the problem analytically with the help of several
approximations.

First, we will assume that the droplet is always perfectly
spherical, with a radius $R$ and a total mass $m$ (\textit{spherical
droplet approximation}, SDA). This is an approximation because
anchoring effects combined with elastic interactions may actually
slightly distort the shape of the droplet into an ellipsoid. The
molecular director configuration within the droplet will be taken to
be axial, namely the director $\n$ has a well defined uniform
direction close to the center of the droplet, while it will be
distorted to some extent close to the surface due to anchoring
\cite{erdmann1990,kralj1992,juodkazis2003}. This allows one to
properly define an average molecular director of the droplet.

Second, for calculating exactly the overall electromagnetic torque
acting on the droplet we would have to determine in all details the
light propagation within the spherical droplet, including all the
birefringence and wave diffraction (Mie scattering) effects: an
exceedingly complex task. Instead, following a common practice in
the literature \cite{friese1998,juodkazis1999,juodkazis2003}, we
will use an approximate expression of the electromagnetic torque
obtained by simply replacing the spherical droplet with a
homogeneous slab of liquid crystal having a thickness equal to the
droplet diameter and the strongly focused light beam with a plane
wave (\textit{planar symmetry approximation}, PSA). This
approximation will tend to become more exact in the limit of large
droplets and weakly-focused light beams.

Third, we will restrict the possible dynamics of the fluid and the
director to either one of the following two approximate models: (i)
the droplet behaves exactly as a rigid body, i.e., rotating only as a
whole and with the director perfectly locked to the fluid
(\textit{rigid body approximation}, RBA); or (ii) the droplet fluid
flow and director are allowed to have different, although uniform,
rotation dynamics but the director field is taken to be perfectly
uniform (\textit{uniform director approximation}, UDA).

Let us now go into the details of the outlined approximations. The
SDA approximation needs no further comments, so we move on to the
calculation of the total electromagnetic torque acting on the
droplet within the PSA approximation.

\subsection{Total external electromagnetic torque}
We assume that a focused light beam passes through a liquid crystal
droplet and that the average molecular director inside the droplet is
oriented perpendicular to the beam axis. We choose a reference system
in which the $z$ axis coincides with the beam axis and the average
molecular director lies in the $xy$ plane. Note that, even if
initially the average director of the droplet will not necessarily lie
in the $xy$ plane, the electromagnetic torque itself will force it
there, in order to align the director to the optical electric field.
So, at steady state, our assumption will be always verified.

In order to calculate the total electromagnetic torque we must use
either Eq.\ (\ref{couplevol}) or Eq.\ (\ref{couplesurf}), with
$\alpha=\elm$ and with expressions (\ref{tauem}) and
(\ref{emstress}) of the torque density and stress tensor,
respectively. The main difficulty is that the field to be used in
the integrals is the total one, including both the external input
field and the diffracted or scattered one. Neglecting the latter
will give a vanishing result. So we need to calculate the
propagation of light in the birefringent droplet. The first
approximation introduced here for this calculation consists of
simply replacing the droplet with a uniform planar slab of nematic
liquid crystal having the same molecular director as the average one
in the droplet. Therefore, all diffraction effects are neglected and
the only optical effects left to be considered are the changes of
polarization due to birefringence (and dichroism) and eventually the
attenuation due to absorption. Moreover, any distortion of the
director configuration induced by light itself is assumed to be
negligible, due to elastic interactions. As a second approximation,
we treat the input light as a monochromatic plane wave propagating
along the slab normal $z$. These two approximations combined are
here named ``planar symmetry approximation'' (PSA). We stress that,
despite its common usage
\cite{friese1998,juodkazis1999,juodkazis2003}, PSA is very rough for
the typical experimental geometry of strongly focused light beams
and rather small droplets. Therefore, we can only anticipate
semi-quantitative accuracy of its predictions. For example, in a
strongly focused beam a large fraction of optical energy is actually
associated with waves propagating obliquely, at a large angle with
respect to $z$, which will see a much reduced birefringence with
respect to the PSA plane wave. At any rate, all the model
inaccuracies associated with the PSA approximation will not be very
different for pure and dye-doped droplets.

The slab thickness is taken equal to the droplet diameter $d=2R$.
The liquid crystal birefringence is denoted as $\Delta{n}=n_e-n_o$,
where $n_o=\text{Re}\left(\sqrt{\varepsilon_{\perp}}\right)$ and
$n_e=\text{Re}\left(\sqrt{\varepsilon_{\perp}+\varepsilon_a}\right)$
are the ordinary and extraordinary refractive indices, respectively.
The absorption coefficient is denoted as $\alpha_0$ (we neglect the
dichroism for simplicity). The input light beam properties are the
total power $P_0$, angular frequency $\omega=2{\pi}c/\lambda$,
vacuum wavenumber $k=2\pi/\lambda$, and a polarization assumed to be
elliptical with its major axis parallel to the $x$ axis and a degree
of ellipticity fixed by the reduced Stokes parameter $s_3$ or
equivalently the ellipsometry angle $\chi$ (in a complex
representation of the input plane wave, their definition is
$s_3=\sin(2\chi)=2\text{Im}(E_xE^*_y)/(|E_x|^2+|E_y|^2)$).

The calculation of the output wave fields emerging at the end of the
slab is lengthy but straightforward, so we skip it. Inserting the
input and output fields in Eq.\ (\ref{couplesurf}) and integrating
(the integration surface $\partial\Vo$ will be given by the two
planes delimiting the slab and corresponding to the input and output
fields; moreover, it is necessary to start the calculation with a
finite wave and then take the plane-wave limit only after having
performed a first integration by parts \cite{humblet43}), we obtain
the following final expression of the external electromagnetic
torque \cite{friese1998}:
\begin{eqnarray}
M^{\elm}_z&=&\frac{P_0}{\omega}\left\{s_3\left[1-e^{-2\alpha_0
R}\cos(\Delta\phi)\right] \right. \nonumber\\ && \!\!\!\!\!\left.
-\left(\sqrt{1-s_3^2}\right)e^{-2\alpha_0
R}\sin(\Delta\phi)\sin2\theta\right\}, \label{torqueslab}
\end{eqnarray}
where $\theta$ is the angle between the director $\n$ and the $x$
axis within the $xy$ plane and $\Delta\phi=2kR\Delta{n}$ is the
total birefringence retardation phase.

It is interesting to note that the two main terms appearing in
Eq.~(\ref{torqueslab}) tend to induce conflicting dynamics. The
first, maximized for a circularly polarized input light
($s_3={\pm}1$) and independent of the director orientation, tends to
induce a constant rotation around the $z$ axis, in a direction fixed
by the sign of $s_3$. The second term, instead, maximized for a
linearly polarized input light ($s_3=0$) and dependent on the
director orientation, tends to align the average molecular director
of the droplet either parallel or perpendicular to the $x$ axis,
i.e., the major axis of the input polarization ellipse, depending on
the birefringence retardation $\Delta\phi$. For small values of
$s_3$ the latter term dominates and there is always an equilibrium
angle $\theta$ at which $M^{\elm}_z$ vanishes. If, instead, the
polarization ellipticity $|s_3|$ is larger than a certain threshold
$s_{3t}=\sin(2\chi_t)$ such that the former term dominates for any
value of $\theta$, the torque $M^{\elm}_z$ cannot vanish and the
droplet must keep rotating (although not uniformly, unless
$s_3=\pm1$). The threshold ellipticity is defined by the following
equation:
\begin{equation}
\frac{s_{3t}}{\sqrt{1-s_{3t}^2}}=\tan(2\chi_t)=\frac{e^{-2\alpha_0
R}\sin(\Delta\phi)}{1-e^{-2\alpha_0 R}\cos(\Delta\phi)}.
\label{threshold}
\end{equation}

It will be useful to consider also the mathematical limit of Eq.\
(\ref{torqueslab}) for $\alpha_0\rightarrow\infty$. This corresponds
to the case in which the torque contribution of the light emerging
from the output plane of the slab vanishes completely, as it occurs
for very large absorption. However, we note that the same
mathematical result is also obtained by taking the average of
Eq.~(\ref{torqueslab}) over a wide range of birefringence
retardations $\Delta\phi$, so that oscillating terms are canceled
out. Such an average may occur as a result of two factors neglected
in our PSA model: (i) oblique propagation of strongly focused light
in the droplet (in our opinion this is the strongest effect),
leading to reduced $\Delta n$ and hence $\Delta\phi$; (ii)
propagation of light off droplet center, leading to an optical path
length that is shorter than $2R$. Whatever the actual cause, in this
limit the electromagnetic torque reduces to the simple expression
\begin{equation}
M^{\elm}_z=M^{\elm}_{z0}=\frac{s_3P_0}{\omega},
\label{torqueslabinputonly}
\end{equation}
corresponding to the total flux of ``spin'' angular momentum
associated with the input light only.

Let us now turn to the droplet dynamics.

\subsection{Droplet dynamics in the rigid body approximation (RBA)}
The RBA approximation can be justified by the fact that the typical
viscosity (Leslie's) coefficients of the nematic liquid crystals (a
typical value is $\gamma_1\approx 100$ cP) are much larger than the
water viscosity ($\eta\approx 1$ cP at room temperature). So any
internal shear or relative rotation of the director with respect to
the droplet fluid will be much slower than the overall droplet
rotation with respect to the surrounding water. Moreover, it is
possible that rigid-body behavior of the droplet (in particular in the
steady-state dynamical regimes) is further enforced by the elastic
interactions in combination with anchoring conditions (this second
effect is especially plausible in the case of imperfect sphericity of
the droplets).

Within the RBA, the fluid velocity in the droplet is given by
\begin{equation}
\mathbf{v}(t)=\Om(t){\times}\rb,
\label{rigidv}
\end{equation}
where $\Om(t)$ is the droplet angular velocity. Moreover, the
molecular director is taken to rotate, everywhere in the droplet, at
the same angular velocity as the fluid, i.e., it satisfies the
following equation
\begin{equation}
\frac{d\n}{dt}=\Om(t){\times}\n(t).
\label{rigidn}
\end{equation}
Since the orientation of the average director of the droplet in the
$xy$ plane is given by the angle $\theta$ introduced in the previous
section, one also has $\Om_z=d\theta/dt$.

As with all rigid bodies, all one needs in order to determine the
droplet rotational dynamics is Eq.\ (\ref{cardinal}) for the angular
momentum rate of change, as applied to the entire droplet volume
$\Vo=\Vo_d$. By introducing Eqs.\ (\ref{rigidv}) and (\ref{rigidn}) in
Eq.\ (\ref{momang}), the total angular momentum of the droplet can be
rewritten as
\begin{equation}
\mathbf{L}=I\Om,
\end{equation}
where $I$ is the total moment of inertia, given by
$I=\int_{\Vo_d}\rho(x^2+y^2)dV=2mR^2/5$, and we have neglected
$I_n$.

The total external torques $\Mb^{\alpha}$ acting on the droplet can
be computed more conveniently using their surface integral
expression, Eq.\ (\ref{couplesurf}). As already discussed in the
previous Section, the surface integral can be actually evaluated on
a surface that lies just outside the liquid crystal droplet
boundary, i.e., within water, thereby making the calculation much
simpler. Let us now consider each of the five contributions
$\alpha=\hp,\el,\vis,\elm,\ph$.

First, owing to the spherical shape of the droplet (within the SDA),
the pressure torque $\Mb^{\hp}$ will vanish identically, as it can
be readily verified by a direct calculation. Since in water there
are no elastic stresses, the elastic torque $\Mb^{\el}$ will also
vanish identically.

The viscous term does not vanish and it can be easily evaluated by
solving the Navier-Stokes equations in water with assigned velocity
on the droplet boundary as given by Eq.\ (\ref{rigidv}) in the
laminar flow limit (and neglecting the effect of the electromagnetic
stresses in water). The result of such calculation is the well known
Stokes formula for the rotational viscous torque acting on a
rotating sphere in a viscous fluid:
\begin{equation}
\Mb^{\vis}=-6\eta\Vo_d\Om,
\label{waterfriction}
\end{equation}
where $\eta$ is the water viscosity coefficient.

The electromagnetic torque $\Mb^{\elm}$ does not vanish. Within the
PSA approximation discussed above, its $z$ component will be given
by Eq.\ (\ref{torqueslab}), while its $x$ and $y$ components will
vanish.

Finally, as already discussed in the previous Section, the
photoinduced torque $\Mb^{\ph}$ should also vanish based on our
theory, as in water there should be no photoinduced stresses.
However, as discussed above, we cannot exclude that a higher-order
theory might predict a nonvanishing $\Mb^{\ph}$ associated to
interfacial effects (a specific possibility for such an effect would
be, for example, a photoinduced discontinuity of the flow velocity
at the boundary between liquid crystal and water). Therefore, we
should consider this possibility in our analysis.

Neglecting inertial terms, equation (\ref{cardinal}) in the RBA
model is then reduced to the following torque balance:
\begin{equation}
\Mb^{\vis}+\Mb^{\elm}+\Mb^{\ph}=0,
\label{RBAmod}
\end{equation}
where the first two torques are given by Eq.~(\ref{waterfriction})
and (\ref{torqueslab}), while the expression of $\Mb^{\ph}$ is
unknown. Equation (\ref{RBAmod}) is actually a first-order
differential equation in the rotational angle $\theta(t)$.

Let us assume initially that $\Mb^{\ph}=0$. As discussed in the
previous section, the steady-state solution of Eq.\ (\ref{RBAmod})
depends on the value of the polarization ellipticity $s_3$ with
respect to the threshold value $s_{3t}$ given in
Eq.~(\ref{threshold}). For $|s_3|<s_{3t}$ the solution is static,
i.e., $\theta(t)=\theta_0$ is constant, while for $|s_3|>s_{3t}$ the
solution is dynamical and corresponds to a generally nonuniform
rotation of the droplet around the $z$ axis. In the circular
polarization limit the rotation becomes uniform.

By a simple integration, it is possible to determine the overall
rotation frequency $f$ of the droplet, which takes the following
expression:
\begin{eqnarray}
f&=&f_0\;\text{Re}\left\{\left[ s_3^2
\left(1-e^{-2\alpha_0 R}\cos\Delta\phi\right)^2 \right.\right. \nonumber\\
&&\;\;\;\;\;\;\;\;\;\;\;\;\left.\left. - (1-s_3^2)e^{-4\alpha_0
R}\sin^2\Delta\phi\right]^{\frac{1}{2}}\right\},
\label{eq_frequenza}
\end{eqnarray}
where
\begin{equation}
f_0=P_0/(16\pi^2\omega\eta R^3)
\label{eq_f0}
\end{equation}
(note that Eq.~(\ref{eq_frequenza}) includes also the stationary
solutions $f=0$, for $|s_3|<s_{3t}$). The highest frequency is
obviously reached for $s_3={\pm}1$, i.e., for circular polarization
of the input light. Note also that if we take the
$\alpha_0\rightarrow\infty$ limit (which may be actually due to all
the factors discussed above and neglected in PSA), we obtain simply
$f=f_0$ instead of Eq.~(\ref{eq_frequenza}).

Let us now consider what should happen instead for $\Mb^{\ph}\neq0$.
As we said, we do not know the actual expression of a nonvanishing
$\Mb^{\ph}$, as this should result from some unknown higher-order
term in the constitutive equations. However, since the photoinduced
torque density $\btau^{\ph}$ acting on the molecular director is
proportional to the electromagnetic one $\btau^{\elm}$, it is
reasonable to expect that also this photoinduced torque $\Mb^{\ph}$
is proportional to $\Mb^{\elm}$. The ratio of the photoinduced to
electromagnetic torque density is $\zeta/\varepsilon_a$, a number
which is of the order of several hundreds. The corresponding ratio
of total external torques is therefore limited by this value,
although it could be smaller.

In the case of circularly polarized input light, the effect of the
photoinduced external torque would be that of inducing a
dye-enhanced droplet rotation, as revealed by a higher rotation
frequency achieved for the same input light power, or a smaller
light power needed to obtain the same rotation frequency when
compared with the undoped case or to what happens when light falling
outside the dye absorption band is used. In the experimental section
we will specifically search for such effects.

\subsection{Uniform director approximation (UDA)}
According to the RBA model presented in the previous subsection, the
rotation speed of a droplet is independent of all photoinduced
effects, unless higher-order interfacial effects should be found to be
significant. This conclusion relies strongly on the assumption that
the droplet rotates effectively like a rigid body. In this subsection
we analyze theoretically a situation in which the director is not
constrained to rotate together with the fluid flow. Such a situation
may occur with appropriate boundary conditions for the director
orientation and fluid motion at the droplet interface. We show that in
this case the velocity field and the director rotation are influenced
essentially by photoinduced effects.

In order to keep the model manageable analytically, we assume here
that the spatial distribution of the director in the droplet is always
approximately uniform (this assumption was unnecessary in the RBA
model). Moreover, we will still consider the fluid motion to coincide
with that of a rigid body.

The dynamics of $\mathbf{v}$ and $\n$ fields will therefore still be
taken to be given by Eqs.\ (\ref{rigidv}) and (\ref{rigidn}), but the
two $\Om$'s entering these equations will be different, in general.
Let us then label $\Om_v$ and $\Om_n$ the angular velocities of the
fluid and director rotations, respectively.

We can find two dynamical equations for $\Om_n$ and $\Om_v$ starting
from Eqs.~(\ref{eqforn}) and (\ref{newton}), respectively, and
following a procedure similar to that used when finding
Eq.~(\ref{cardinal}). First, we integrate both sides of
Eq.~(\ref{eqforn}) over the whole droplet volume $\Vo_d$. Second, we
multiply both sides of Eq.~(\ref{newton}) vectorially by $\rb$ and
integrate over the whole volume $\Vo_d$. Owing to Eq.~(\ref{tautrel}),
we can then write the two resulting equations in the following form:
\begin{eqnarray}
I_n\Vo_d\frac{d\mathbf{\Omega}_n}{dt}&=&\Tb^{\tot} \nonumber\\
I\frac{d\mathbf{\Omega}_v}{dt}&=&\Mb^{\tot}-\Tb^{\tot},
\label{eqsnv}
\end{eqnarray}
where the total external torque
$\Mb^{\tot}=\sum_{\alpha}\Mb^{\alpha}$ is still defined by
Eq.~(\ref{couplesurf}) and it is therefore identical to that used in
the RBA model, and we have also introduced the total
\textit{internal torque} $\Tb^{\tot}=\sum_{\alpha}\Tb^{\alpha}$
exchanged between director and velocity degrees of freedom. More
precisely, for each kind of interaction
$\alpha=\hp,\el,\vis,\elm,\ph$, the torque $\mathbf{T}^{\alpha}$ is
simply defined as the volume integral of the torque density
$\btau^{\alpha}$ over the whole droplet.

Equations (\ref{eqsnv}) highlight the coupling between the droplet
fluid rotation and the director dynamics inside the droplet. One
recovers the RBA model when the internal torque $\Tb^{\tot}$ is very
rigid and acts as a constraint that locks the director to the fluid.

Since $I_n \approx 0$ with excellent accuracy, from the first of
Eqs.~(\ref{eqsnv}) one finds $\mathbf{T}^{\tot}\approx0$. Then the
second of Eqs.~(\ref{eqsnv}) becomes identical to that of the RBA.
This result, however, should not be taken as saying that within the
UDA the droplet always behaves exactly as in the RBA, as the
external torques $\Mb^{\alpha}$, and in particular the
electromagnetic one, can be affected by the director orientation in
the droplet, which will not be the same as in RBA. In particular, as
we will see, the photoinduced effects will be present even in the
first-order theory, in which they only appear via the internal
torque $\Tb^{\ph}$.

Let us now calculate the internal torques $\mathbf{T}^{\alpha}$
using the explicit constitutive dependence of the corresponding
director torque densities $\btau^{\alpha}$. The hydrostatic pressure
contribution vanishes identically. For a perfectly uniform director
the elastic torque density $\btau^{\el}$ and therefore the total
torque $\mathbf{T}^{\el}$ also vanish identically. However it is not
obvious that one can truly neglect the elastic torque density, as it
is just this torque that keeps the director approximately uniform.
If all other torque densities are always uniform then the director
will remain uniform ``spontaneously'', and the elastic torque
density will indeed vanish. Whenever the other torque densities are
nonuniform (in our case, for example, the electromagnetic torque
density is not very uniform in large droplets), their effect will
give rise to a nonvanishing elastic torque density which balances
them, in order to constrain the director to remain approximately
uniform. At any rate, we assume here that the total elastic torque,
obtained by integrating this elastic torque density over the whole
droplet, remains always negligible. Further work will be needed to
assess the validity of this assumption.

Let us now consider the other interactions. The viscous torque under
our assumptions is easily calculated and is given by
\begin{equation}
\mathbf{T}^{\vis}=-\gamma_1\Vo_d(\Om_n-\Om_v),
\end{equation}
where $\gamma_1$ is the orientational viscosity coefficient
\cite{degennes2ed,chandrasekhar2ed}.

As for the electromagnetic term, under our hypothesis of uniform
director (UDA) and spherical droplet (SDA) of constant density, it can
be shown that the following identity holds with high accuracy:
\begin{equation}
\mathbf{T}^{\elm}=\Mb^{\elm}.
\label{emcoupleid}
\end{equation}
The proof of this identity is reported in the Appendix. Thereby, the
calculation of $\Tb^{\elm}=\Mb^{\elm}$ can be based on the PSA
approximation, and its explicit expression is given by
Eq.~(\ref{torqueslab}).

Finally, the photoinduced term (in first-order theory, the only one
we consider in this section) is simply related to the
electromagnetic one when the latter is caused by an optical wave
having a wavelength within the dye absorption band. In this case,
since $\btau^{\ph}=(\zeta/\epsilon_a)\btau^{\elm}$, the photoinduced
internal torque will be exactly given by
\begin{equation}
\mathbf{T}^{\ph}=\frac{\zeta}{\varepsilon_a}\Mb^{\elm}.
\label{dyecouple}
\end{equation}
However, it must be kept in mind that this strict relationship
between the photoinduced and the electromagnetic torques is only
valid when the eletromagnetic field is that of an optical wave
absorbed by the dye. It is instead possible to separately tune the
electromagnetic and photoinduced torques by adding a second wave
whose wavelength is out of the dye absorption band, or alternatively
by adding static electromagnetic fields. In these cases,
Eq.~(\ref{dyecouple}) will only apply to the contribution of the
wave having a frequency within the dye absorption band.

By using these results, Eqs.~(\ref{eqsnv}) can be rewritten in the
following dynamical equations, which define our UDA model of the
droplet dynamics:
\begin{eqnarray}
I_n\Vo_d\frac{d\Om_n}{dt}\!&=&\!\!-\gamma_1\Vo_d(\Om_n-\Om_v)
+\Mb^{\elm}+\mathbf{T}^{\ph}\approx 0 ,\nonumber\\
I\frac{d\Om_v}{dt}\!&=&\!\!-6\Vo_d\eta\Om_v+\gamma_1\Vo_d(\Om_n
-\Om_v)-\mathbf{T}^{\ph}, \label{eqsnv1}
\end{eqnarray}
where in the second equation we have cancelled the two electromagnetic
terms $\Mb^{\elm}$ and $-\mathbf{T}^{\elm}$ as they are exactly
opposite to each other.

Equations (\ref{eqsnv1}) show explicitly the internal nature of the
photoinduced effects (according to first-order theory), with the
``action'' and ``reaction'' torques appearing respectively in the
former and latter equation (or vice versa). In contrast, the
electromagnetic torque is external (the ``reaction'' term acts on
the electromagnetic field) and is applied on the director only, in
the first place. It is only via the viscous internal coupling
between the director and the fluid motion (and possibly also the
elastic one, which we neglected) that the electromagnetic torque
finally drives also the droplet fluid rotation.

Let us now study the dynamics predicted by Eqs.~(\ref{eqsnv1}) in a
couple of interesting examples.

Let us assume first that a single circularly polarized laser beam
travelling along the $z$ axis drives the droplet rotation. The
director will spontaneously orient in the $xy$ plane. Then, the
electromagnetic torque becomes independent of the director
orientation within the plane, so that its rotation may become
uniform at steady state. In this case, Eqs.~(\ref{eqsnv1}) are
readily solved and give
\begin{eqnarray}
\Om_{nz} & = & \frac{M^{\elm}_z}{\Vo_d}\left(
\frac{1}{6\eta}+\frac{1+\zeta/\varepsilon_a}{\gamma_1}\right),\nonumber\\
\Om_{vz} & = & \frac{M^{\elm}_z}{6\eta\Vo_d},
\label{steadyrot}
\end{eqnarray}
where $M^{em}_z$ can be approximately calculated using
Eq.~(\ref{torqueslab}). These results show that the droplet fluid and
the director do not rotate at the same rate. For $\zeta=0$ (no
photoinduced effect), the difference in angular velocity is very small
as the viscous coefficient $\gamma_1$ is about two orders of magnitude
larger than $\eta$. In contrast, in the presence of photoinduced
effects, the ratio $\zeta/\varepsilon_a$ can be larger than a few
hundreds \cite{janossy91mclc,janossy92}, so that a significant
difference in the two angular velocities should become possible.

As a second example, let us assume that there are two linearly
polarized optical waves travelling along $z$ and driving the droplet
dynamics: one is assumed to be polarized along $x$ and to have a
wavelength falling outside the dye absorption band; the second is
instead assumed to be polarized at an angle $\Psi$ with the $x$ axis
and to have a wavelength that is within the dye absorption band.
Having two waves with only one being absorbed by the dye, the
electromagnetic and photoinduced torques can be adjusted
independently to each other.

Being linearly polarized, the two waves will induce torques that
tend to align the droplet director parallel or perpendicular to the
polarization direction of the respective wave. So we may assume that
at steady state the director will acquire a fixed orientation along
some intermediate direction $\theta$ with $0<\theta<\Psi$.
Therefore, at steady state one may assume $\Om_n=0$ and the
electromagnetic and photoinduced torques to be constant in time. By
solving Eqs.~(\ref{eqsnv1}) with these assumptions, one finds that
the equilibrium orientation of the director is actually fixed by the
following balance:
\begin{equation}
M^{\elm}_z+\frac{6\eta}{\gamma_1+6\eta}T^{\ph}_z=0,
\end{equation}
while the \textit{droplet fluid rotates constantly} with the following
angular velocity:
\begin{equation}
\Omega_v = \frac{M^{\elm}_z}{6\eta \Vo_d} = -\frac{T^{\ph}_z}{(\gamma_1+6\eta)\Vo_d}.
\end{equation}
This result is rather counterintuitive: a continuous droplet
rotation is induced by two linearly polarized optical waves having
different planes of polarization. This could not occur without the
presence of photoinduced effects, i.e., based only on the
electromagnetic torque (it must not be forgotten that the two waves
have different wavelengths, so their superposition is not coherent).
We see, then, that the photoinduced effects associated with the dye
can give rise to highly nontrivial effects in a liquid crystal
droplet, as long as the director can be decoupled from the fluid
motion.

\section{Experiment}
We prepared emulsions of pure and dye-doped liquid crystal (LC) in
bidistilled water. By properly choosing the relative quantities of
liquid crystal and water, we could obtain relatively stable
emulsions containing many liquid crystal droplets having micrometric
size, most of them in the 1-20 $\mu$m diameter range. Most
experiments were performed using the commercial liquid crystal
mixture E63 (see Ref.\ \cite{jakli91} for its composition), provided
by Merck, Darmstadt, Germany. This material is convenient for the
wide temperature range of its nematic phase (from -30$^{\circ}$C to
82$^{\circ}$C). A few experiments were performed using the liquid
crystal 4-cyano-4'-pentyl-biphenyl (5CB). The dye used in doped
materials was the 1,8-dihydroxy 4,5-diamino 2,7-diisopentyl
anthraquinone (HK271, provided by Nematel, Mainz, Germany), known as
one of the most effective dyes in the photoinduced effects. We
prepared dye-liquid crystal solutions at a concentration of 2\% in
weight, leading to torque enhancement ratios $\zeta/\varepsilon_a$
of several hundreds \cite{janossy91mclc,janossy92}.

\begin{figure}[h]
\includegraphics[angle=0, width=0.48\textwidth]{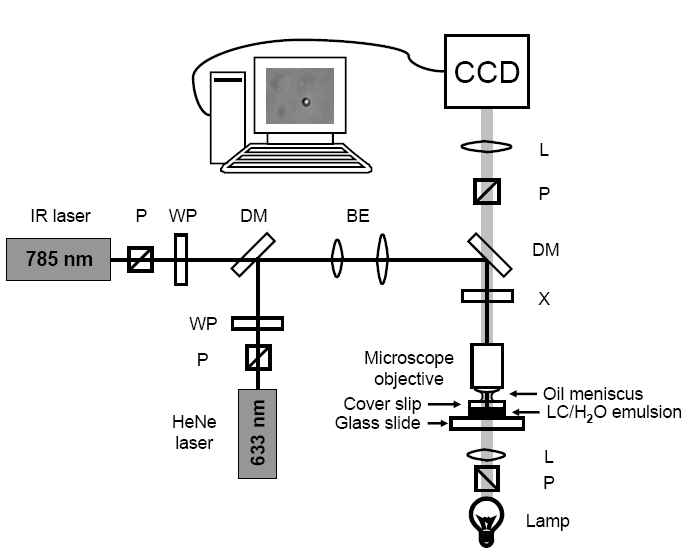}
\caption{Experimental setup. Legend: P polarizer, WP half-wave plate, DM
dichroic mirror, L lens, CCD - ccd camera, BE beam expander for
objective back aperture overfilling, X either retardation wave-plate
(for the appropriate laser wavelength) or optically active plate (see
text) for rotating the linear polarizations of the two wavelengths by
a different amount, depending on the specific experiment.}
\label{fig_apparato}
\end{figure}

For performing the optical rotation experiments, a small volume of
emulsion was placed on a microscope glass slide and covered with a
thin (thickness of about 0.18 mm) glass coverslip, thus forming a
thin cell (open on the sides). A 100${\times}$ oil-immersion
microscope objective (Carl Zeiss, NA=1.25) was used both to image
the droplets on a CCD camera (using lamp illumination from below the
cell, in combination with other optics) and to focus the input laser
beams that were used to trap the droplets and induce their rotation,
as discussed in the following (see Fig.~\ref{fig_apparato}). The
objective was attached, via an oil thin film, to the glass
coverslip. When needed, the microscopic imaging could be made in the
crossed or nearly-crossed polarizers geometry, in order to visualize
the droplet birefringence. CCD images and movies (acquisition rate
of 50 frame/s) could also be recorded on a PC, for subsequent
frame-by-frame analysis.

The setup for the optical trapping and manipulation is a
dual-wavelength optical-tweezers apparatus (see
Fig.~\ref{fig_apparato}). Our setup differs from usual optical
tweezers arrangement in that two laser beams, respectively generated
by a diode-pumped solid state laser ($\lambda=$785 nm, subsequently
called ``IR'' beam) and a He-Ne laser ($\lambda$=633 nm), are
brought to a common focus at the specimen plane of the microscope
objective. Both beams were able to achieve trapping of droplets for
diameters in the 2-12 $\mu$m range, as proved by a sudden stop (or
confinement) of the droplet Brownian motion after trapping. The
beam-waist (Airy disk) radius at focus is estimated to be about 0.4
$\mu$m, in the adopted objective-overfilling configuration.
Actually, due to the thickness of the glass coverslip, our optical
trap center (roughly corresponding to the laser beam waist) could
not be located right in the middle of the cell but was close to the
coverslip. This fact might have led to some contacting of the
droplets with the glass, particularly in the case of the largest
droplets. In such cases the droplets may have experienced a somewhat
stronger friction, and therefore our results on the rotational speed
could be biased. However, any systematic effect will be identical
for pure and dye-doped droplets and obviously independent of the
light wavelength used for inducing the rotation. So, our comparison
will remain valid. Moreover, the agreement between theory and
experiments will show that these possible systematic effects are
certainly small, at least for not too large droplets.

The common focus of the two lasers allowed us to switch the
controlling beam from one wavelength to the other (e.g., to switch
the He-Ne beam on and the IR off, or vice versa) without changing
the trapped liquid crystal droplet. Since the He-Ne wavelength is
close to the maximum of the dye absorption band, while the IR
wavelength falls completely outside the absorption band, this
corresponds to adding or removing the photoinduced effects, so as to
allow for a direct comparison of the behavior obtained in the two
cases with the same droplet.

Besides for trapping, the two beams were also used to induce the
electromagnetic and photoinduced torques acting on the droplet and
driving its reorientation or continuous rotation. By inserting along
the beam (just before the microscope objective) a suitable
birefringent wave plate, we could control the polarization of each
beam (separately). In particular, we have used linear polarizations
with an adjustable polarization plane and elliptical polarizations
with an adjustable degree of ellipticity $s_3=\sin(2\chi)$. We could
also obtain certain specific polarization combinations of both beams
together, such as one linear and the other circular or elliptical.
By using a suitably dispersive optically-active plate (home-built
using a chiral-doped randomly oriented liquid crystal cell) we could
also obtain two linearly polarized beams with two different
polarization planes, forming an angle of $\Psi\approx40^{\circ}$.

Micron-sized LC droplets are known to show either one of two
possible director spatial distribution
\cite{erdmann1990,kralj1992,juodkazis2003}: axial (or bipolar),
already discussed in the previous section, and radial, which has
full spherical symmetry and a hedgehog defect at the droplet center.
The images of axial and radial droplets under a polarizing
microscope can be almost identical. However, we could identify axial
droplets by checking that their image changed if the microscope
polarizers were rotated, or by looking directly at their dynamical
behavior under the laser beam, as only axial droplets could be
readily set in rotation if illuminated with circular polarizations
or reoriented using linear polarizations. Radial droplets could not
be rotated at all (because all the optical torques vanish, due to
radial symmetry, if absorption is neglected). About 80-90\% of the
droplets of our emulsions were found to be axial \cite{footnote4}.
In the following we will only refer to them.

Using linearly polarized light, we could easily fix the average
director orientation of our trapped axial droplets. By looking at the
microscopic image pattern under crossed
polarizers~\cite{juodkazis1999apl, sandomirski2004}, we verified that
the director orientation was indeed parallel to the polarization
plane. By slowly rotating the polarization by means of a half-wave
plate, the droplet alignment followed the polarization. These
experiments could be done both with pure liquid crystal droplets and
with dye-doped ones, the latter both with IR and He-Ne beams, without
much difference.

\begin{figure}[h]
\includegraphics[angle=0, width=0.48\textwidth]{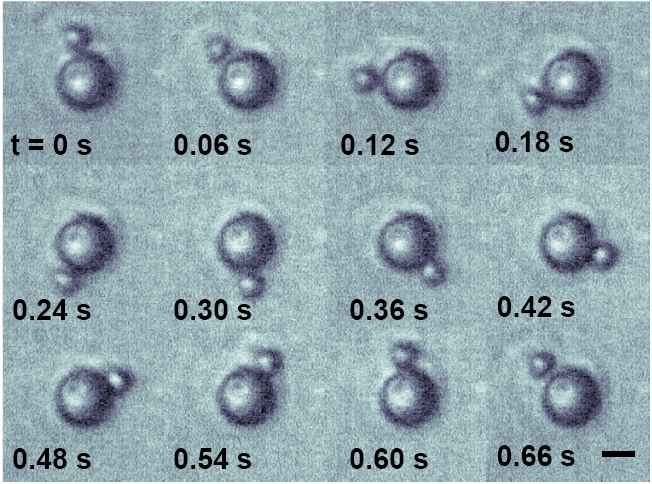}
\caption{Sequence of photograms showing the light-induced rotation
of a LC droplet in time, as highlighted by the revolution of a dragged
small object. Scale-bar: 1$\mu$m.}
\label{fig_rotazione}
\end{figure}

Using elliptically polarized beams and axial droplets, we could
perform the analogue of the classic Beth's
experiment~\cite{beth1936}, i.e., set the trapped droplets in
continuous rotation by the transfer of angular momentum with light.
Depending on the droplet size and on the laser beam power and
ellipticity, the rotation could range from very slow (periods of
several seconds) to very fast (down to the millisecond range). We
used two different methods for measuring the spinning of droplets by
analyzing the frames of a droplet rotation movie. First, using the
crossed-polarizers geometry we could image the birefringence
rotation, corresponding to the director motion, as a periodic
pattern modulation in time. For the smallest droplets, we could
observe a significant modulation of light intensity during the
rotation even without the analyzer, presumably due to the
anisotropic scattering cross-section of the droplets. Second, we
could also measure the droplet fluid rotation (i.e., regardless of
director rotation) when a smaller satellite droplet (or another
small object) happened to get trapped close to the rotating droplet
and was thereby dragged around, as in the case shown in
Fig.~\ref{fig_rotazione}. The rotation speed measured in these two
ways (when the satellite was sufficiently small and close to the
rotating droplet) were always found to be the same.

In order to check if the photoinduced effects give rise to a
rotation speed enhancement, on each droplet we measured the rotation
frequency induced by the He-Ne and the IR beams as a function of
input beam polarization ellipticity. We repeated this for many
different droplets having a range of diameters. We also measured the
rotation frequencies induced in pure (undoped) liquid crystal
droplets, although of course in this case the comparison could not
be done with exactly the same droplet sizes. In all these
measurements, the two laser beams were adjusted for having a roughly
equal angular momentum flux, as given by
Eq.~(\ref{torqueslabinputonly}). In particular, the light power
measured after the microscope objective was about 4.1 mW for the
He-Ne beam and 2.8 mW for the IR in most data shown. These values
correspond, for a given polarization ellipticity $s_3$, to the same
angular momentum flux to within 15\%
($M^{\elm}_{z0}=(1.3{\pm}0.1){\times}10^{-18}$ Nm for the case of
circular polarization). However we also investigated the power
dependence of the rotation frequency in some droplets.

\begin{figure}[h]
\includegraphics[angle=0, width=0.48\textwidth]{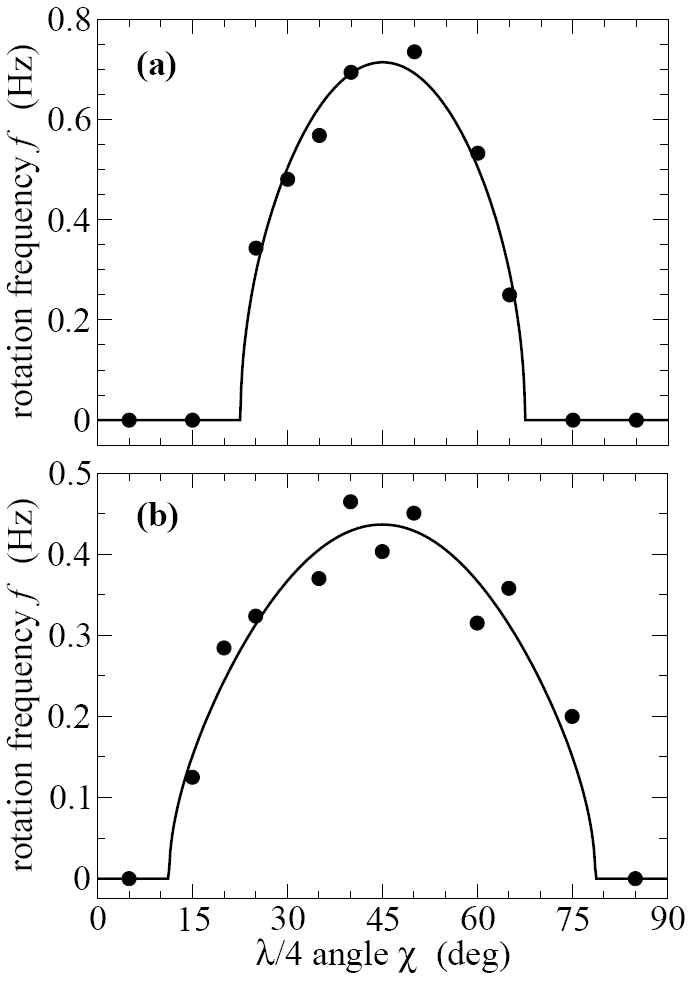}
\caption{Droplet rotation frequency $f$ versus input light
ellipticity angle $\chi$, for a dye-doped LC droplet having a radius
of 2.4 $\mu$m, using the IR (a) or He-Ne (b) laser beams, with a
power of 2.8 mW and 4.1 mW, respectively. The dots are the measured
values and the solid line is the theoretical fit based on Eq.\
(\ref{eq_frequenza}). The threshold ellipticity $\chi_t$ is obtained
from the fit from the two symmetrical points at which the solid line
crosses zero.} \label{fig_campana}
\end{figure}

An example of the measured rotational frequencies of a given droplet
versus light polarization ellipticity (given by the angle $\chi$) is
shown in Fig.~\ref{fig_campana}. We actually measured this
dependence for many other droplets of different sizes, made of both
pure and dye-doped LC. Each of these measurements was then fitted by
means of Eq.~(\ref{eq_frequenza}). In these fits the radius $R$ was
fixed at the value determined by analyzing the droplet microscopic
picture. The absorbance at 633 nm was instead measured separately on
a bulk sample (we obtained $\alpha_0=(1.0\pm0.2){\times}10^{3}$
cm$^{-1}$) and then kept fixed to this value in all fits. The
constants $f_0$ and $\Delta\phi$ were adjusted for best fit. $f_0$
was found to be roughly consistent (to within a factor two in most
cases) with its predicted value, as given by Eq.~(\ref{eq_f0})
(using the known or measured values of the laser power and
frequency, water viscosity and droplet radius). From the best-fit
value of $\Delta\phi$, we could also estimate the birefringence
$\Delta{n}$, which was almost always found to fall in the range
0.11-0.13 (average 0.12), with a few droplets giving a value of 0.10
and 0.14. These values are inconsistent with the known refractive
index difference of the bulk material ($\Delta{n}=0.2273$ at 589 nm,
Merck data sheet). We ascribe this discrepancy to the strong
approximations associated with the PSA model (in particular to
neglecting the contribution of obliquely propagating waves in the
focused beam), as a strong depression of the optical anisotropy due
to confinement effects is not plausible for micrometric droplets. It
is apparent from Fig.~\ref{fig_campana} that the rotational behavior
of a dye-doped LC droplet under IR or He-Ne laser beams is not
identical. However, the difference is well explained by the
different wavelength and absorbance in Eq.~(\ref{eq_frequenza}),
while \textit{no particular rotation enhancement is seen} in the
He-Ne case, where photoinduced dye effects should take place.

\begin{figure}[h]
\includegraphics[angle=0, width=0.48\textwidth]{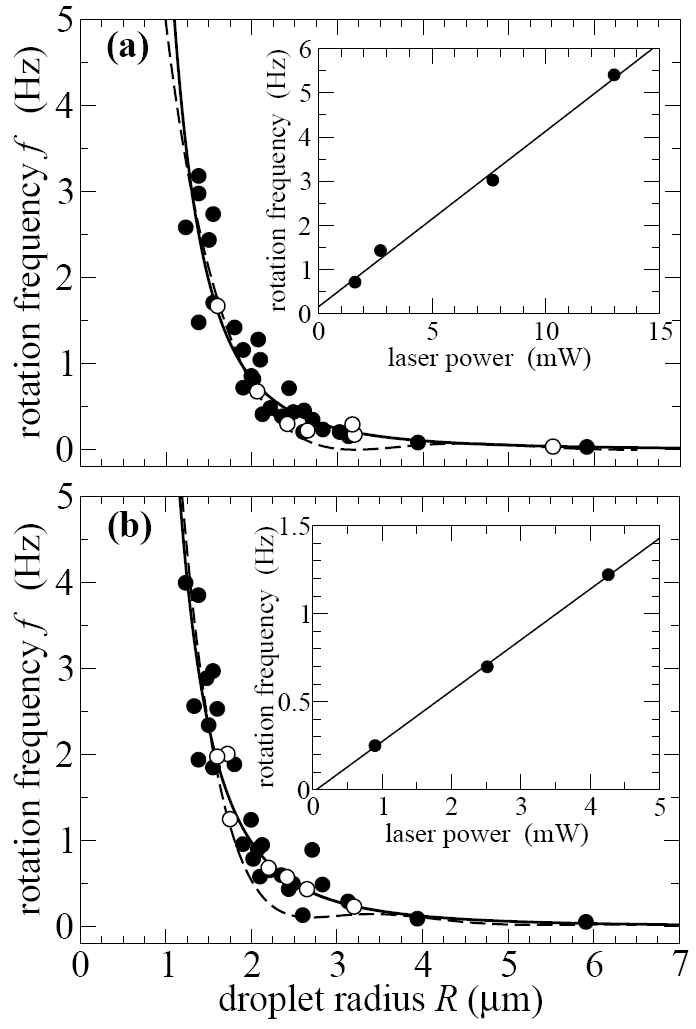}
\caption{Frequency $f$ of the droplet rotation induced by circularly
polarized light. The main panels show the dependence on droplet
radius for a fixed laser power (power on the sample: 4.1 mW for the
He-Ne and 2.8 mW for the IR beams), while the insets show the
dependence on laser power for a fixed droplet radius of 1.8 $\mu$m.
Panel (a) refers to rotations induced by the IR laser, panel (b) to
the He-Ne case. Filled (open) circles refer to droplets made of
dye-doped (pure) liquid crystal. Solid and dashed lines are the
theoretical predictions obtained as explained in the text.}
\label{fig_fvsr}
\end{figure}

In Fig.~\ref{fig_fvsr} we show the rotation frequency observed for
several pure LC (open symbols) and dye-doped LC (filled symbols)
droplets of different sizes using circularly polarized light. Panels
(a) and (b) refer to rotations induced by the IR and He-Ne beam,
respectively. The insets show the corresponding (linear) dependence
on laser power for a fixed droplet (incidentally, this linear
behavior supports our assumption that the laser light induces no
significant distortion of the director configuration in the droplet,
in the power range used in our experiments).

Again, a first conclusion that can be immediately drawn from this
figure is that \textit{no significant rotation speed enhancement takes
place} in the dye-doped LC case with the He-Ne beam as compared both
to the IR beam case and to the pure LC case. Should the photoinduced
torque be external, one would have expected a rotation speed
enhancement by a factor of the order of $\zeta/\epsilon_a$, i.e., of
several hundreds.

In Fig.~\ref{fig_fvsr}, the dashed lines are the predictions of
Eq.~(\ref{eq_frequenza}) after adjusting the laser power $P$ for
best-fit to data. The birefringence was kept fixed to the average
value $\Delta{n}=0.12$ obtained from the measurements discussed
above (and confirmed by the threshold ellipticity data, as discussed
below), but increasing its value led to worse fits (in particular,
$\Delta{n}=0.23$ leads to very bad fits). From Fig.~\ref{fig_fvsr},
it is seen that the agreement between data and theory is reasonable,
although there is a statistically significant discrepancy. In
particular, the data do not show at all the oscillations predicted
by Eq.~(\ref{eq_frequenza}) (of course, small residual oscillations
might be hidden in the noise). Moreover, for both lasers the
best-fit values of the power were found to be about a factor two
smaller than the actual measured values (assuming a water viscosity
$\eta=1$ cP, corresponding to the room temperature of
20$^{\circ}$C). The solid lines in Fig.~\ref{fig_fvsr} correspond
instead to the simpler theory $f(R)=f_0(R)$, as given by
Eq.~(\ref{eq_f0}), obtained in the limit
$\alpha_0\rightarrow\infty$. In this case, \textit{no adjustable
parameter was used}, i.e., the values of the laser power and of the
water viscosity are both fixed to the known values ($\eta=1$ cP).
Nevertheless, it is seen that the agreement is even better and that
the same theory (taking into account the difference in light power
and frequency) explains all the data, i.e., both the dye-doped case
with He-Ne beam, when there is significant absorption, and the IR
beam or pure LC cases, when there is no absorption (incidentally,
this agreement shows that any systematic effect due to the droplet
closeness to the glass wall is essentially negligible, except
perhaps for the largest droplets). This better agreement of the
simplified theory obtained for $\alpha_0\rightarrow\infty$ clearly
cannot be truly ascribed to absorption (negligible in the IR and
pure droplets case). It is instead the likely result of light
diffraction, oblique propagation and other effects neglected in the
simple PSA calculation, which may all contribute to averaging out
the oscillations due to the outgoing light, as discussed previously.
This view is also confirmed by the fact that previous works on
transparent liquid crystal droplets reported similar observations
\cite{juodkazis1999,juodkazis2003}.

\begin{figure}[h]
\includegraphics[angle=0, width=0.48\textwidth]{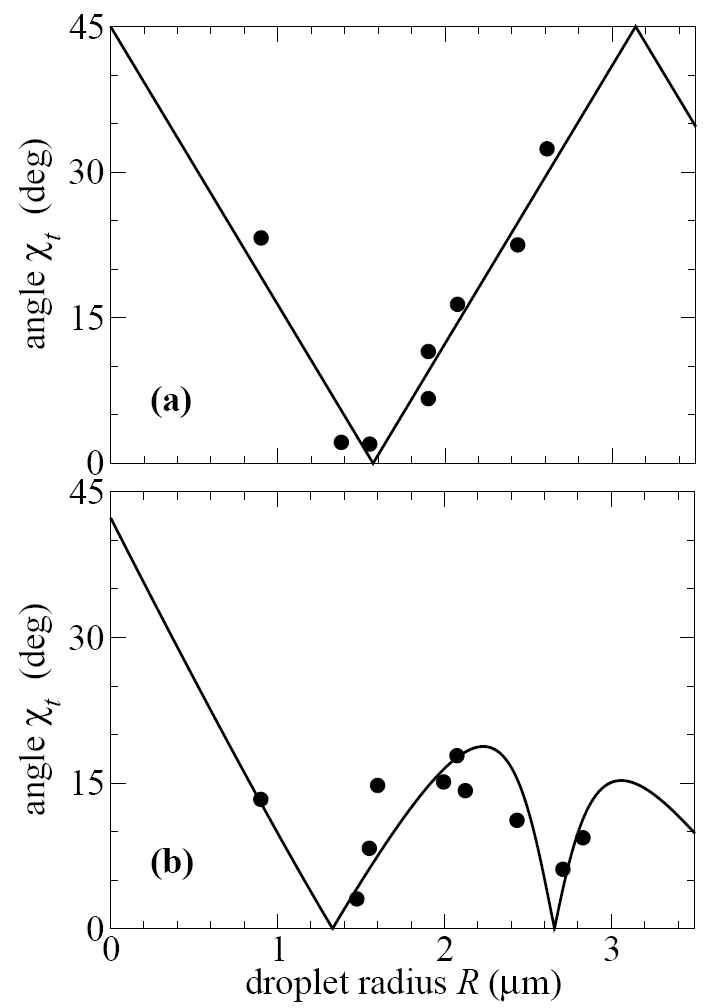}
\caption{Threshold ellipticity for droplet rotation $\chi_t$ as a
function of droplet radius. Data points are actually obtained from
the fits described in the text and in the caption of Fig.\
\ref{fig_campana}. The solid line is from Eq.~(\ref{threshold}).
Panel (a) refers to the IR case, panel (b) to the He-Ne. All data
are for droplet of dye-doped liquid crystal.} \label{fig_chitvsr}
\end{figure}

Let us now consider the behavior of the threshold ellipticity
$\chi_t$ for droplet rotation versus the droplet radius $R$. For
each droplet and wavelength, $\chi_t$ was obtained from the best
fits performed on the measured rotation frequency versus light
ellipticity (such as that shown in Fig.~\ref{fig_campana}). In
particular, $\chi_t$ is entirely determined by the best-fit value of
$\Delta\phi$, via Eq.~(\ref{threshold}). The resulting data are
shown in Fig.~\ref{fig_chitvsr}, together with the predictions
obtained using Eq.~(\ref{threshold}) combined with
$\Delta\phi=2kR\Delta{n}$ and adjusting the birefringence $\Delta n$
for best fit. It is seen that the theory agrees reasonably well with
the experiment. The best fit is for $\Delta n\approx0.12$ for both
laser wavelengths, confirming the average value obtained before.
Moreover, for the He-Ne case the effect of light absorption is
clearly seen in the data.

It must be noted that Eq.~(\ref{threshold}) is based on the PSA
model \textit{including} the effect of the light emerging from the
droplet. Indeed, taking the $\alpha_0\rightarrow\infty$ limit leads
to $\chi_t=0$ for all droplet sizes, which is inconsistent with our
data. We do not know exactly why the PSA-neglected effects do not
affect the threshold ellipticity as much as the rotation frequency
(apart from the depression of $\Delta{n}$), but this is what we
actually find and what was also reported in previous works
\cite{juodkazis1999,juodkazis2003}. At any rate, we stress that all
inaccuracies of PSA theory do not affect our comparison between pure
and dye-doped droplets or between IR and He-Ne beams.

Finally, we tried to find evidences of a decoupling between the
droplet fluid and the average director degrees of freedom. In
particular, we searched for the effects predicted by the UDA model to
occur when the He-Ne and IR beams are simultaneously impinging on the
droplet, both linearly polarized with the two polarization planes
forming an angle of about 40$^{\circ}$. In order to visualize the droplet
fluid rotation independently of the director, we imaged the dynamics
of droplets having small satellites (such as smaller droplets or other
particles in the suspension), so that the droplet surface flow would
be highlighted by the revolution of these dragged objects. However,
despite many efforts, we could not find any sign of a steady-state
decoupling between the director and the fluid motion in the droplets
(except, perhaps, for some complex transient effects) and, therefore,
of a corresponding continuous fluid rotation of the droplet. This
probably indicates that elastic effects associated with the nonuniform
director distribution and a not perfectly spherical shape of the
droplet (effects neglected in the UDA model), concur to keeping the
orientational and flow degrees of freedom locked to each other.

\section{Conclusions}
In summary, we have studied for the first time the dynamical
behavior of droplets of dye-doped nematic liquid crystal trapped in
water by a strongly focussed laser beam and set in rotation by the
optical torques generated by the same beam. We searched for
evidences of a dye-induced enhancement of the droplet rotation speed
associated with photoinduced effects, but no enhancement was found.
This null result is in accordance with the leading first-order
theory of these photoinduced effects \cite{janossy94,marrucci97pre}
and directly proves that the dye-enhanced optical torque is not
associated with an exchange of angular momentum with light or other
external degrees of freedom, but that it must instead be associated
with a fully internal exchange of angular momentum between the
molecular orientation and fluid flow degrees of freedom. The latter
exchange is correctly described only by assuming the existence of a
dye-induced stress tensor acting on the fluid flow in the
illuminated region, as first proposed in Ref.\ \cite{marrucci97pre}.
This photoinduced stress tensor embodies the internal ``reaction''
to the photoinduced torque acting on the molecular director. Our
null result is therefore also a direct proof of the actual existence
of this stress tensor.

The interplay between fluid flow and director dynamics within the
droplet is also predicted by our models to give rise to very
peculiar effects, such as a continuous droplet fluid rotation
induced by two linearly polarized optical waves having different
polarization planes. However, the occurrence of these phenomena
requires to unlock the constraint existing between the droplet fluid
and the molecular director. This constraint is partly due to viscous
forces, but it may also be due to elastic interactions combined with
anchoring forces. Experimentally, we could not reach a situation in
which this constraint was effectively broken, so we could not
demonstrate the predicted effect.

\acknowledgments
We thank R. Eidenschink of Nematel for providing HK271. This work was
supported by the Fondo per gli Investimenti della Ricerca di Base
(FIRB) of the Italian Ministero dell'Istuzione, dell'Universit\`{a} e
della Ricerca. I. J. acknowledges the support from the Hungarian
Research Grants OTKA T-037275 and NKFP-128/6.

\appendix*
\section{Proof of the electromagnetic torque identity}
To prove the validity of identity (\ref{emcoupleid}), we start from
Eq.\ (\ref{couplevol}) as applied to the case $\alpha=\elm$, which can
be rewritten as follows:
\begin{equation}
M^{\elm}_i-T^{\elm}_i=I_i=\int_{\Vo}\epsilon_{ijh}x_j\partial_k\te^{\elm}_{kh}.
\label{nullint}
\end{equation}
We must therefore prove that the integral $I_i$ vanishes identically
within the SDA and UDA approximations.

Let us consider the first term in Eq.\ (\ref{emstress}) of the
electromagnetic stress tensor. This is an isotropic pressure-like
term, i.e., of the form $p\delta_{hk}$. For such a term, the integral
(\ref{nullint}) can be recast into a surface integral as shown in the
following:
\begin{eqnarray}
I^{(1)}_i&=&\int_{\Vo}\epsilon_{ijh}x_j\partial_kp\delta_{hk}dV=
\int_{\Vo}\epsilon_{ijh}x_j\partial_hpdV \nonumber\\
&=&\int_{\Vo}\epsilon_{ijh}\partial_h(x_jp)dV=
\oint_{\partial\Vo}\epsilon_{ijh}x_jpdA_h. \nonumber
\end{eqnarray}
Within the SDA, the surface element normal $dA_h$ and the position
vector $x_j$ will be parallel to each other, and therefore their cross
product (as expressed by their product times the antisymmetric
Levi-Civita tensor) vanishes identically.

Let us now consider the second term in Eq.\ (\ref{emstress}) of the
stress tensor. Omitting the time-average for brevity, we must consider
the following integral
\begin{eqnarray}
I^{(2)}_i&=&\!\!\int_{\Vo}\epsilon_{ijh}x_j\partial_k(D_kE_h)dV=
\int_{\Vo}\epsilon_{ijh}x_jD_k\partial_kE_hdV\nonumber\\
&=&\!\!\int_{\Vo}\epsilon_{ijh}x_jD_k\partial_hE_kdV-
\int_{\Vo}\epsilon_{ijh}x_jD_k\epsilon_{khl}\frac{\partial{B_l}}{\partial{t}}dV\nonumber\\
&\!\!\!\!=&\!\!\int_{\Vo}\epsilon_{ijh}x_j\varepsilon_{kl}E_l\partial_hE_kdV-
\int_{\Vo}\epsilon_{ijh}x_jD_k\epsilon_{khl}\frac{\partial{B_l}}{\partial{t}}dV,\nonumber
\end{eqnarray}
where we have used the two Maxwell equations $\partial_kD_k=0$ and
$\partial_kE_h-\partial_hE_k=-\epsilon_{khl}\partial{B_l}/\partial{t}$
and introduced the dielectric tensor to express $\mathbf{D}$ in terms
of $\E$. The dielectric tensor is always symmetric for a permutation
of its indices. Moreover, within UDA and constant density
approximations, the dielectric tensor is also uniform within the
droplet. Exploiting these two properties, we obtain
\begin{eqnarray}
I^{(2)}_i&\!\!\!\!=&\!\!\!\!\frac{1}{2}\!\int_{\Vo}\!\epsilon_{ijh}x_j\partial_h(\varepsilon_{kl}E_lE_k)dV
\!-\!\!\int_{\Vo}\!\epsilon_{ijh}x_jD_k\epsilon_{khl}\frac{\partial{B_l}}{\partial{t}}dV\nonumber\\
&=&-\int_{\Vo}\epsilon_{ijh}x_jD_k\epsilon_{khl}\frac{\partial{B_l}}{\partial{t}}dV,\nonumber
\end{eqnarray}
where the first integral vanishes identically (within SDA) as it has
taken a pressure-like form.

The third stress-tensor term, in the magnetic field, can be treated
analogously to the second, obtaining the following expression:
\begin{displaymath}
I^{(3)}_i=\int_{\Vo}\epsilon_{ijh}x_jB_k\epsilon_{khl}\frac{\partial{D_l}}{\partial{t}}dV.
\end{displaymath}

Summing up the three contributions, the whole integral
$I_i=I^{(1)}_i+I^{(2)}_i+I^{(3)}_i$ is reduced to the following:
\begin{eqnarray}
I_i&=&\int_{\Vo}\epsilon_{ijh}x_j\frac{\partial}{\partial{t}}(\epsilon_{khl}D_lB_k)dV\nonumber\\
&=&\int_{\Vo}\left[\rb{\times}\frac{\partial}{\partial{t}}(\mathbf{D}{\times}\mathbf{B})\right]_idV.
\end{eqnarray}

This expression (which is a part of the electromagnetic torque due
the so-called Abraham force \cite{landau}, which is completed if one
considers also the effect of the neglected electromagnetic force
mentioned in \cite{footnote2}) corresponds to a true residual
electromagnetic torque which does not vanish in general. However,
after optical-cycle averaging, its order of magnitude can be
estimated as follows:
\begin{equation}
I_i \simeq |\mathbf{D}||\E|\Vo_d\left(\frac{\Delta{t}}{\tau}\right)
\simeq T^{\elm}_i\left(\frac{\Delta{t}}{\tau}\right),
\end{equation}
where $\tau$ is a characteristic time of the droplet dynamics and
$\Delta{t}\approx{R}/c$ is the droplet crossing time of light. In our
case $\Delta{t}/\tau\approx10^{-13}$, making this contribution totally
negligible with respect to $T^{\elm}_i$ and therefore the identity
(\ref{emcoupleid}) valid to high accuracy.

%\bibliography{droplets}

\begin{thebibliography}{26}
\bibitem{footnote1}
More generally one must use the local order-parameter traceless
tensor $\mathsf{S}_{ij}(\rb)$. This can be particularly important in
the smallest droplets, where significant order parameter depressions
can be induced by the strong director gradients.
\bibitem{footnote2}
The divergence of the stress tensor actually does not provide the
complete electromagnetic force in electro-dynamical systems. An
additional force of the form
$-(\partial\mathbf{S}/\partial{t})/c^2$, where $\mathbf{S}$ is the
Poynting vector of the optical wave, should be included (this is
part of the so-called Abraham force), as discussed in Ref.\
\protect\cite{landau}. However, after optical-cycle averaging this
last term is negligible whenever the matter dynamics is very slow
with respect to the droplet crossing time by light, as in our case
(see the Appendix for a fuller explanation).
\bibitem{footnote3}
In equation (\ref{momang}), a contribution to the angular momentum
density associated with the electromagnetic field itself, of the
form $\rb{\times}\mathbf{S}/c^2$, has been neglected for the same
reasons as those discussed in \protect\cite{footnote2}.
\bibitem{footnote4}
When doing experiments with 5CB droplets, it was relatively easy to
heat them above the clearing temperature (nematic to isotropic
transition), either by laser-induced heating of the single droplet
or by heating the whole sample. In any case, radial droplets usually
transformed into axial after being heated above the clearing point
and then left relaxing back to the nematic phase. This showed that
axial alignment is probably the most stable one.
\end{thebibliography}

\newpage

\newpage

\newpage

\newpage

\newpage

\end{document}